\documentclass{iau} 
\usepackage{graphicx}

\title[EoR Extra-galactic Foregrounds] 
{EoR Foregrounds:\\ the Faint Extragalactic Radio Sky}

\author[I. Prandoni]   
{Isabella Prandoni}

\affiliation{Institute of Radioastronomy, INAF, \\ Via P. Gobetti 101,
IT-40129, Bologna, Italy \\ email: {\tt prandoni@ira.inaf.it}}

\pubyear{2018}
\volume{333}  
\jname{Peering towards Cosmic Dawn}
\editors{Vibor Jeli\'c \& Thijs van der Hulst, eds.}
\begin{document}

\maketitle

\begin{abstract}
A wealth of new data from upgraded and new radio interferometers are rapidly improving and transforming our understanding of the faint extra-galactic radio sky. Indeed the mounting statistics at sub-mJy and $\mu$Jy flux levels is finally allowing us to get stringent observational constraints on  the faint radio population and on the  modeling of its the various components. In this paper I will provide a brief overview of the latest results in areas that are potentially important for an accurate treatment of extra-galactic foregrounds in experiments designed to probe the Epoch of Reionization. 
\end{abstract}

\firstsection 
\section{Introduction}
The  redshifted  21-cm  hyperfine  transition  line  of  neutral  hydrogen is one of the most promising methods to directly probe the  intergalactic  medium  during  the earliest epochs of galaxy formation, the so-called Cosmic dawn (CD) and Epoch of Reionization (EoR). Statistical or direct detection of this cosmological signal is indeed one of the prime goals of  new-generation radio facilities operating at low frequency ($<200$ MHz), like e.g. LOFAR$^1$, MWA$^2$ and SKA-LOW$^3$, and the scope of a number of dedicated experiments  (e.g. PAPER$^4$, HERA$^5$). \footnotetext{\\$^1$www.lofar.org \\ $^2$www.mwatelescope.org \\ $^3$www.skatelescope.org \\ $^4$eor.berkeley.edu \\ $^5$reionization.org}\\
One of the main challenges of EoR experiments is, however, the strong contamination from systematic effects (ionospheric distortions, telescope response, etc.) and astrophysical foregrounds (Galactic and extra-galactic), a few  orders of magnitude brighter than the cosmological signal (e.g. \cite[Jeli\'c et al. 2008]{Jelic2008}; \cite[Bernardi et al. 2009]{Bernardi2009}; \cite[Jeli\'c et al. 2010]{Jelic2010}; \cite[Zaroubi et al. 2012]{Zaroubi2012}; \cite[Pober et al. 2013]{Pober2013}; \cite[Dillon et al. 2014]{Dillon2014}; \cite[Parsons et al. 2014]{Parsons2014}; \cite[Chapman et al. 2015]{Chapman2015}). 

The astrophysical foregrounds can be treated in different ways. All methods rely on the basic fact that, differently from the HI-redshifted cosmological signal, foreground  sources  are  continuum  emitters,  and  therefore  have  a  smooth  spectrum over a small bandwidth. The foregrounds can be directly  {\it removed} from the data by fitting each detected source with some spectral model (see \cite[Chapman et al. 2015]{Chapman2015} for a review). 
Alternatively the foregrounds can be {\it avoided}. A useful diagnostic tool in this regard is the 2D cylindrical power spectrum, that bins modes in the plane of the sky, k$_{\perp}$, and modes parallel to the line of sight, k$_{||}$. Spectral smoothness confines the majority of foreground power to large parallel scales (low k$_{||}$). The EoR signal can then be effectively searched for by averaging only over high-k$_{||}$ modes (the so-called {\it EoR window}), in which the foreground power is low (e.g.  \cite[Parsons et al. 2012]{Parsons2012}; \cite[Jacobs et al. 2015]{Jacobs2015}). The foregrounds can be otherwise {\it suppressed}, by assigning appropriate weights to modes in which the foregrounds are expected to dominate (e.g., \cite[Liu \& Tegmark 2011]{Liu2011}). Finally, a mixed approach can be used (e.g. \cite[Trott et al. 2016]{Trott2016}). It is clear that an accurate modeling of EoR foregrounds is crucial for the implementation of effective mitigation algorithms.  

A wealth of new data from upgraded and new radio interferometers are rapidly improving and transforming our understanding of the faint extra-galactic radio sky. In this paper I will give an overview of the latest results, focusing, when possible, on low frequency surveys, which are the most relevant for EoR experiments. In addition I will focus on star-forming galaxies (SFG) and (low-power) Active Galactic Nuclei (AGN), the two point-source populations dominating the faint extra-galactic radio sky. Other populations (like, e.g., diffuse sources associated with clusters -- radio halos and relics; AGN remnants, etc.) are much less relevant from a statistical point of view, and will be here neglected (but see Sect.~\ref{sec-misc}). 

Three areas of investigation stand out when attempting  accurate modeling  of the SFG and AGN populations contributing to the faint radio sky: we need a good knowledge of 1) their physical and evolutionary properties at radio band (that can be conveniently integrated into a modeling of flux-dependent radio source counts);  2) their sky distribution (i.e. their clustering properties), and 3) their radio Spectral Energy Distributions (SED).  Indeed EoR extra-galactic foreground simulations have to make assumptions on all these three modeling aspects. Source counts are often implemented as single power-laws (see e.g.  \cite[Trott et al. 2016]{Trott2016}), as well as radio spectra; the source spatial distribution is either assumed to be Poissonian (i.e. statistically uniform; see e.g.  \cite[Trott et al. 2016]{Trott2016}), or very simple clustering laws are implemented (e.g. \cite[Jeli\'c 2010]{Jelic2010};  \cite[Liu \& Tegmark 2011]{Liu2011}). As I will show in the following, too simplistic assumptions are not adequate to accurately reproduce the complex mixture of radio source populations, that are detected at sub-mJy and $\mu$Jy flux densities. This may be potentially harmful for  EoR experiments. For instance, \cite[Murray et al. (2017; see also this Volume)]{Murray2017} have demonstrated that neglecting source clustering may result in underestimating the foreground power and overestimating the size of the EoR window, potentially leading to a false detection of signal. 

Another potentially important area of investigation when dealing with EoR experiments regards the foreground polarization properties.  Indeed the leakage of polarized emission into total intensity, in presence of inaccurate calibration of the off-axis instrumental polarization, may produce a spurious signal, that can mimic the cosmological one (\cite[Jeli\'c et al. 2014]{Jelic2014}).  Reliable modeling of polarized (extra-galactic and Galactic) foregrounds can provide reliable predictions of such an effect (see e.g. the recent work by \cite[Nunhokee et al. 2017]{Nunhokee2017}). 

\section{Radio Source Counts and the Quest for New Evolutionary Models}\label{sec-counts}

It is well established that the sub-mJy population has a composite nature. While star formation processes in galaxies  dominate at $\mu$Jy levels (e.g. \cite[Seymour et al. 2008]{Seymour2008}), Radio Loud (RL) AGNs are the dominant component at flux densities $>0.5$ mJy (e.g. \cite[Mignano et al. 2008]{Mignano2008}). Recently, it has been shown that a significant fraction of the sources detected  below $\sim 100$ $\mu$Jy are associated with Seyfert galaxies or QSO. Most of these sources do not display large-scale radio jets and lobes, typical of RL AGN, and for this reason they are often referred to as RQ AGN.  The origin of their radio emission is hotly debated: it may come from star formation in the host galaxy (\cite[Padovani et al. 2011]{Padovani2011}; \cite[Kimball et al. 2011]{Kimball2011}; \cite[Condon et al. 2013]{Condon2013}; \cite[Bonzini et al. 2013; 2015]{Bonzini2013}) or from AGN cores (\cite[Maini et al. 2016]{Maini2016}; \cite[Herrera-Ruiz et al. 2016]{HR2016}; \cite[2017]{HR2017}; see also \cite[White et al. 2015]{White2015}; \cite[2017]{White2017};). The most likely  scenario is  that RQ AGN are composite systems where star formation and AGN radio emission can coexist (e.g. \cite[Del Vecchio et al. 2017]{Delvecchio2017}). Depending on how RQ AGN are treated (as a separate class or as part of the star forming radio population), two or three-component models can be developed to describe the sub-mJy population. It has proven to be very difficult, however, to discriminate between these two approaches as, until very recently, the deepest surveys suffered from limited statistics at the relevant fluxes ($\sim 10 - 500$  $\mu$Jy). As illustrated below, thanks to the growing number of  wide-field (degree-scale) surveys sensitive to flux densities $\lesssim 500$ $\mu$Jy, such models can be now put to the test. This is possible only at GHz frequencies (1-3 GHz) though, because low frequency surveys have not yet reached the needed depth. Nevertheless we expect this will change very soon with the deep, wide-area surveys ongoing at LOFAR.

\begin{figure}[t]
\begin{center}
\includegraphics[width=3.7in,angle=-90]{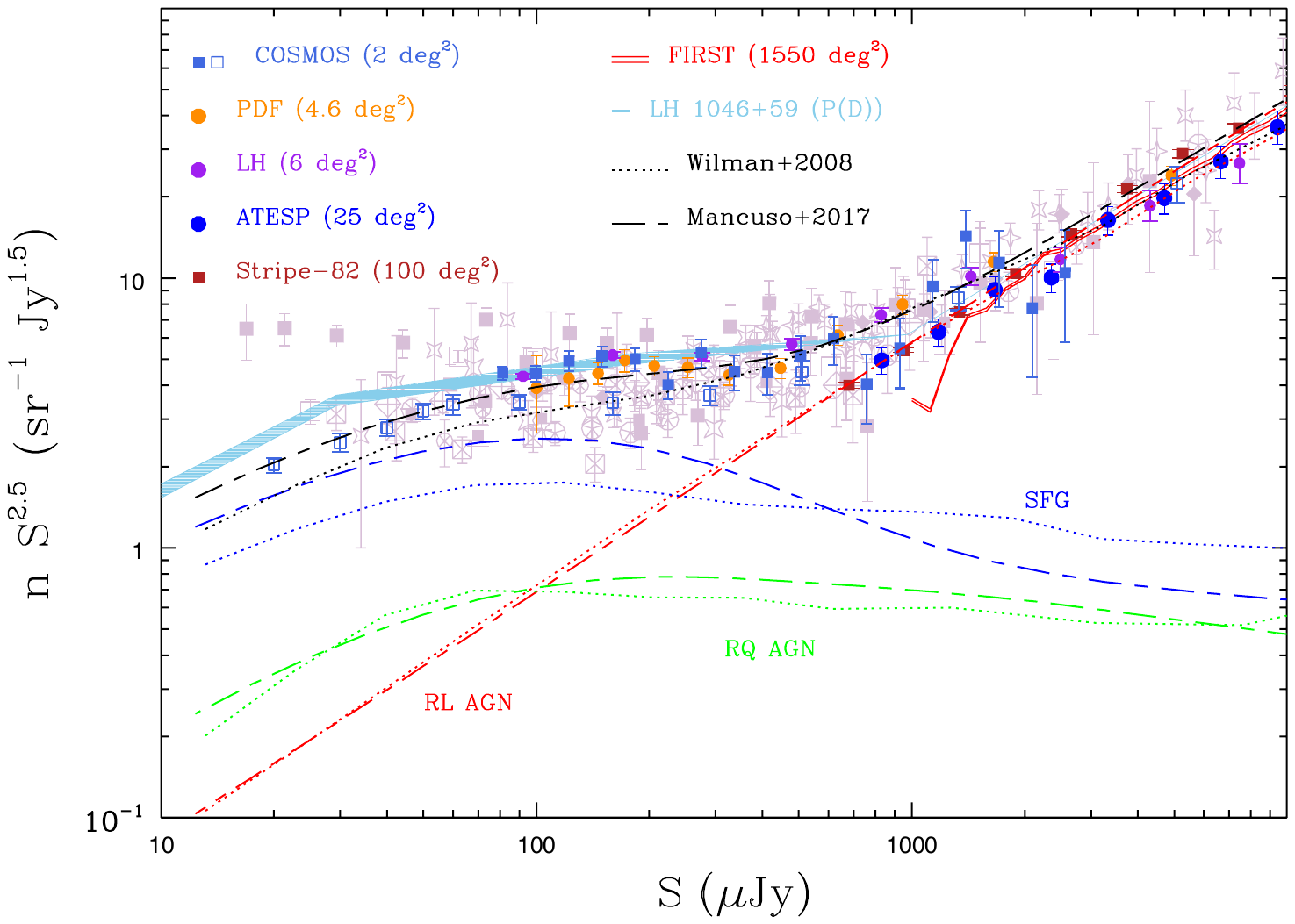}
\vspace*{-0.2 cm}
 \caption{Normalized 1.4 GHz differential source counts. Large-scale surveys are indicated by colored symbols: COSMOS field (2 deg$^2$;  \cite[Bondi et al. 2008]{Bondi2008}; \cite[Smolcic et al. 2017b]{Smolcic2017b}; light blue squares); Phoenix Deep Field (PDF; 4.6 deg$^2$;  \cite[Hopkins et al. 2003]{Hopkins2003}; orange filled circles); the Lockman Hole region (6 deg$^2$; \cite[Prandoni et al. 2018]{Prandoni2018}; magenta filled circles); the ATESP survey (25 deg$^2$; \cite[Prandoni et al. 2001]{Prandoni2001}; blue filled squares); the Stripe-82 region (100 deg$^2$;  \cite[Heywood et al. 2016]{Heywood2016}; brown filled squares); the FIRST survey (1550 deg$^2$;  \cite[White et al. 1997]{White1997}; red lines). Also highlighted is the result of the $P(D)$ analysis performed by  \cite[Vernstrom et al. (2014]{Vernstrom2014}, cyan shaded area). In this case the counts are rescaled from 3 to 1.4 GHz by assuming $\alpha$=-0.7. Small-scale surveys are shown in background (grey symbols; see \cite[Prandoni et al. 2018]{Prandoni2018} and references therein). Vertical bars represent Poissonian errors on the normalized counts. The dotted and dashed lines represent predicted counts from respectively the S3-SEX simulation (\cite[Wilman et al. 2008]{Wilman2008}; \cite[2010]{Wilman2010}), and  the M17 model. Different colours indicate different populations: SFG (blue), RL AGN (red), RQ AGN (green), and their sum (black). We caveat that the M17 RQ AGN track shown here includes radio silent (RS) AGN (see text and Fig.~\ref{fig-counts2} for more details). 
   \label{fig-counts}}
\end{center}
\end{figure}

\begin{figure}[t]
\begin{center}
\includegraphics[width=3.7in,angle=-90]{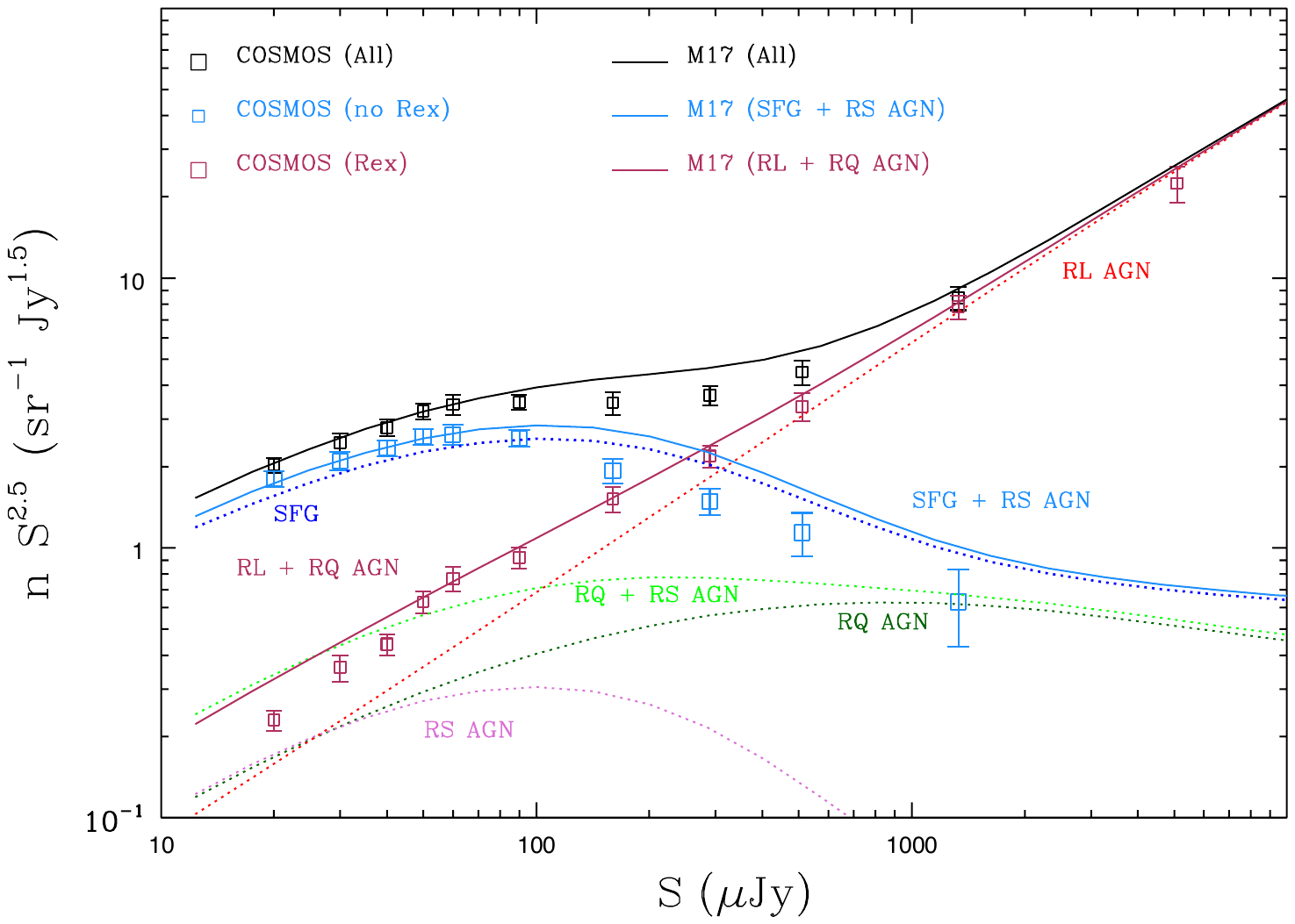}
 \caption{Normalized 1.4 GHz differential source counts. 
Comparison between M17 models (solid lines) and the VLA-COSMOS 3 GHz Large Project dataset (squares; \cite[Smolcic et al. 2017b]{Smolcic2017b}). Colors refer to different populations:  AGN--dominated radio sources (red), SF--dominated radio sources (blue) and their sum  (black). These two source classes are defined as {\it radio excess} and {\it no radio excess} sources by \cite[Smolcic et al. (2017b)]{Smolcic2017b}; in M17 model they are obtained by summing RL and (AGN-dominated) RQ AGN on one side and SFG and RS (SF-dominated RQ) AGN on the other. The match between data and models is very good. Also shown (dotted lines) are the other M17 model components: blue, red and green colors refer to the same classes as shown in Fig.~\ref{fig-counts} (i.e. SFG, RL AGN and the sum of AGN- and SF-dominated RQ AGN. This latter class is here explicitely referred to as RQ + RS AGN.  Dark green and magenta colors show the relative contribution of these two sub-components as a function of flux density.
   \label{fig-counts2}}
\end{center}
\end{figure}

Figure~\ref{fig-counts} shows an updated  collection of source counts obtained at 1.4 GHz. This figure illustrates very well the long-standing issue of the large scatter present at flux densities $\lesssim 1$ mJy, that makes difficult to get useful modeling constraints.  Survey systematics  can play a role in producing this scatter  (see e.g. \cite[Condon et al. 2012]{Condon2012}; \cite[Prandoni et al. 2018]{Prandoni2018}), as well as field-to-field source density fluctuations (cosmic variance; see \cite[Heywood et al. 2013]{Heywood2013}). Indeed when we limit our source counts analysis to the largest deep fields (colored symbols in Fig.~\ref{fig-counts}), the scatter below $1$ mJy gets reduced with respect to when small-scale surveys ($<<1$ deg$^2$) are included (see grey symbols). If we limit our analysis to the deepest degree-scale surveys (COSMOS, Lockman Hole, PDF; but see also the $P(D)$ analysis by \cite[Vernstrom et al. 2014]{Vernstrom2014}), it is clear that the measured source counts show a tendency to be  higher than the ones predicted by the S3-SEX simulation  (black dotted line; \cite[Wilman et al. 2008]{Wilman2008}; \cite[2010]{Wilman2010}) below $\sim 500$ $\mu$Jy, indicating the need to revise such simulation. A full discussion of the various existing models developed to describe the sub-mJy and $\mu$Jy radio populations is beyond the scope of this paper, and I refer to  \cite[Padovani (2016)]{Padovani2016}, for a very recent comprehensive review. Here I will only present some of the latest results, published last year. In particular I will highlight the results obtained in the framework of the VLA-COSMOS 3 GHz Large Project (\cite[Smolcic et al. 2017a]{Smolcic2017a}), which represents the deepest degree-scale survey available to date, and I will compare such results with the novel modeling proposed by \cite[Mancuso et al. (2017, hereafter M17)]{Mancuso2017}, who exploited a model-independent approach to compute star formation rate (SFR) functions, AGN duty cycles, and the conditional probability of a star-forming galaxy to host an AGN with given bolometric luminosity. 

As shown in Fig.~\ref{fig-counts}, the M17 model provides a better description of the overall observed source counts below $\sim 500$ $\mu$Jy (black dashed line).   Both \cite[Wilman et al. (2008]{Wilman2008}; \cite[2010]{Wilman2010}) and M17 consider a three-component modeling (SFG, RL and RQ AGN; see blue, red and green lines in Fig.~\ref{fig-counts}). A direct comparison shows that the main difference is in the SFG component, that is characterized by a steeper evolution in M17. 

While a good description of total source counts is important, more informative is the ability to accurately reproduce the various components of the sub-mJy population. As shown by M17, their predictions are in very good agreement with the observed fractions of SFG, RQ and RL AGN in a number of deep radio samples, where the  same three-component source classification scheme was applied (see Fig.~7 of M17). 

It is worth noting that usually the distinction between SFG and RQ AGN only reflects the multi--band properties of the radio source host galaxy, and the RQ AGN class may include either AGN-- or SF--dominated  radio sources. 
\cite[Smolcic et al (2017b)]{Smolcic2017b} 
introduced a classification based on the so-called {\it radio excess} (Rex; see \cite[Del Vecchio et al. 2017]{Delvecchio2017} for more details). This classification scheme better reflects the radio properties of the source, as it distinguishes between AGN-- and SF--dominated radio sources, independently from the properties of their host galaxies.  This distinction is also present in M17 models, where RQ AGN can be divided in two sub-components depending on the origin of source radio emission: those dominated by AGN activity, and those dominated by SF activity (called {\it radio silent} or RS AGN in M17). This allows a direct comparison with the \cite[Smolcic et al. (2017)]{Smolcic2017b} dataset. The results of this comparison are shown in Fig.~\ref{fig-counts2}. It is interesting to note that the models (solid lines) can reproduce the data (squares) also under this classification scheme (see caption of Fig.~\ref{fig-counts2} for more details). 

It is fair to mention that other models have been proposed. Some of such models include a negative evolution of the IR/radio correlation for SFG, of the form of  $\sim$(1+z)$^{-k}$ (see e.g. \cite[Novak et al. 2017]{Novak2017} and references therein). As discussed by  \cite[Bonato et al. (2017)]{Bonato2017} this evolution should be mild  ($k \lesssim 0.16$ at 1.4 GHz), otherwise it would produce too high source counts, inconsistent with the observations. A negative evolution was recently reported also at low frequency with LOFAR ($k=0.22\pm 0.05$ at 150 MHz; \cite[Calistro Rivera et al. 2017]{Calistro2017}).

\section{New Insights on Low-frequency Radio Spectra}\label{sec-spectra}

The low-frequency radio spectra of SFG and AGN are dominated by synchrotron emission, commonly described by a single power-law function $S \sim \nu^{\alpha}$, where typically  $\alpha \sim -0.7/-0.8$. Nevertheless, there are processes that can change the shape of the spectra (free-free absorption, synchrotron self-absorption, spectral ageing, etc.). In addition the relative brightness of the different components in both SFG (starburst core, disk) and AGN (core, jets, hotspots) may significantly alter the observed integrated radio SED (see e.g. the recent resolved studies of \cite[Kapinska et al. 2017]{Kapinska2017} for NGC~253 and of \cite[McKean et al. 2016]{McKean2016} for Cyg~A). Indeed ongoing multifrequency spectral studies, including MWA and LOFAR, are rapidly transforming our knowledge of radio SED, showing that radio source spectra can be very complicated. 

As far as faint radio sources are concerned, very interesting results were reported by \cite[Calistro Rivera et al. (2017)]{Calistro2017}. Based on deep  (rms down to $\sim 120-150$ $\mu$Jy/b) LOFAR observations of the Bootes field, they found significant differences in the spectral curvature between SFG and AGN. The radio spectra of SF galaxies exhibit a weak but statistically significant flattening below 300 MHz; AGN radio SED, on the other hand, tend to become steeper.  More quantitatively, the median spectral index value of SFG changes from $\alpha^{1400}_{325} = -0.74_{-0.41}^{+0.27}$ to  $\alpha^{325}_{150} = -0.63_{-0.49}^{+0.57}$; for AGN changes from $\alpha^{1400}_{325} = -0.56_{-0.35}^{+0.42}$ to  $\alpha^{325}_{150} = -0.80_{-0.69}^{+0.55}$. Despite the large scatters, the difference between the low- and high-frequency distributions are highly significant, at a level greater than 99.99\% (KM two-sample test). No evolution of the spectral curvature as a function of redshift is found for SFG or AGN,  but the AGN steepening seems to be luminosity dependent: the spectral curvature $\rightarrow 0$ for the  AGN with brightest torus luminosities.  As hotspots can be the dominant source in total flux density in luminous sources, this could suggest that these frequencies are dominated by hotspot emission. The steepening at low frequencies observed for the weaker AGN sources could be explained  by steep-spectrum components dominating in this frequency regime, while flat-spectrum components become relatively more important at higher frequencies, causing the spectral index to flatten (see \cite[Calistro Rivera et al. 2017]{Calistro2017} for a full discussion). These findings clearly indicate that different power-law slopes should be assumed for AGN and SFG, when modeling the faint radio sky at frequencies relevant for EoR experiments. 

\section{The Clustering Properties of Faint Radio Sources}\label{sec-clustering} 

Any reliable modeling of the faint radio sky should include a realistic source spatial distribution. Clustering studies in the radio domain allow to get a direct view of the spatial distribution of radio-selected samples. The angular clustering properties of radio sources have been derived down to mJy flux levels using the all-sky NVSS and the 1550 deg$^2$ FIRST surveys  (\cite[Overzier et al. 2003]{Overzier2003}). Deeper studies were recently undertaken in the  2 $deg^2$ COSMOS field (\cite[Magliocchetti et al. 2017]{Magliocchetti2017}; \cite[Hale et al. 2018]{Hale2018}).  Reaching $\mu$Jy flux levels allows to  extensively probe low luminosity AGN and SFG (that dominate the faint radio sky). In addition, very well studied fields like COSMOS offer  the advantage of having virtually complete source classification and (photometric) redshift information, allowing to trace both redshift and source type dependences. Thanks to the unprecedented statistics of the 3~GHz COSMOS sample, Hale et al. (2018) were able to separate AGN and SFG in sub-classes. Restricting the analysis to $z<1$, they confirmed   that AGN tend to be more highly clustered than SFG, with estimated biases of $b=2.1 \pm 0.2$ and $b=1.5_{-0.2}^{+0.1}$ respectively. In addition they found that low-accretion rate AGN are more clustered ($b = 2.9 \pm 0.3$) than high-accretion rate AGN ($b = 1.8_{-0.5}^{+0.4}$), suggesting that low-accretion rate AGN reside in higher mass haloes. The clustering of radio-selected SFG, on the other hand, appears to have little dependence on SFR. Hints of an evolution  with redshift are also present, even though wider-area (and possibly deeper) surveys are needed to fully constrain how bias scales with radio source properties and how it evolves with redshift. Ongoing and forthcoming surveys with LOFAR, ASKAP and MeerKAT will play an important role in this respect.

\section{Polarization and Diffuse Extra-galactic Emission}\label{sec-misc}
 
At 1.4 GHz the all-sky catalogue by \cite[Taylor et al. (2009)]{Taylor2009} provides an important benchmark for the polarization (and rotation measure, RM) properties of Galactic and extra-galactic radio sources. Wide-area polarization information  at low frequency is instead quite limited. The all-sky LOFAR Two-meter Sky Survey (LOTSS) polarization survey, currently ongoing, will close this gap, providing a direct determination of the low-frequency polarization properties of radio sources down to 1 mJy rms at 150 MHz (see \cite[van Eck et al. 2018]{vanEck2018} for first results). 

At GHz frequencies, deeper (sub-mJy) polarization studies are available in a number of cosmological fields (e.g. the GOODS-N, \cite[Owen \& Rudnick 2014]{Owen2014}; and the ATLAS fields, \cite[Hales et al. 2014]{Hales2014}), but at the current sensitivities disk galaxies (SFG) can be hardly probed in polarization. The COSMOS field is being targeted  in full polarization at 1.4 GHz with the Jansky VLA, as part of a very ambitious project: the CHILES Continuum Polarization (CHILES Con Pol) survey. When complete, this survey is expected to reach an unprecedented sensitivity of 500 nJy per 4 arcsecond FWHM beam. The MIGHTEE project at MeerKAT  will increase the surveyed area in deep fields by one order of magnitude.

On a longer timescale, SKA  will completely transform our understanding of extra-galactic magnetic fields, either through very dense RM all-sky and deep surveys, or the direct search of diffuse synchrotron emission associated with cosmic web filaments (see e.g. \cite[Vazza et al. 2015]{Vazza2015}).  If actually detected, such diffuse extra-galactic radio emission could represent a novel EoR foreground that needs to be properly modeled and included in mitigation algorithms.

\section{Summary }\label{sec-summary}

This paper provides a brief overview of our current understanding of the faint extra-galactic radio sky, focusing on aspects that are potentially important for an accurate treatment of extra-galactic foregrounds in EoR experiments. 
Indeed mounting statistics at sub-mJy and $\mu$Jy flux levels is providing stringent observational constraints on  the faint radio population and on the  modeling of its main components (SFG, RL and RQ AGN). This review presents some of the most recent results, focusing on three main areas: 1) radio-continuum source number counts and related modeling; 2) low-frequency radio SEDs; 3) source clustering properties. 
The emerging picture is rather complex, with various classes of sources displaying a range of  physical and evolutionary properties. This complexity may need to be properly accounted for when modeling EoR foregrounds, to get effective mitigation algorithms.  Another potentially important area of investigation when dealing with EoR experiments regards  foreground polarization properties. Our current knowledge of the extragalactic polarized sky is rather limited, but SKA will be a game changer in this respect.  

\begin{acknowledgment}
IP thanks Gianni Bernardi and Matt Jarvis for useful discussions, and Claudia Mancuso for providing the evolutionary tracks shown in Figures~1 and 2, which refer to the model discussed in Mancuso et al. (2017). IP acknowledges the support of the Ministry of Foreign Affairs and International Cooperation, Directorate General for the Country Promotion (Bilateral Grant Agreement ZA14GR02 - Mapping the Universe on the Pathway to SKA). This publication has received funding from the European Union's Horizon 2020 research and innovation programme under grant agreement No 730562 [RadioNet].
\end{acknowledgment}

\end{document}